\documentclass[lettersize,journal]{IEEEtran}
\usepackage{amsmath,amsfonts}
\usepackage{algorithmic}
\usepackage{algorithm}
\usepackage{array}
\usepackage[caption=false,font=normalsize,labelfont=sf,textfont=sf]{subfig}
\usepackage{textcomp}
\usepackage{stfloats}
\usepackage{url}
\usepackage{verbatim}
\usepackage{graphicx}
\usepackage{cite}
\begin{document}

\title{Decade-Bandwidth RF-Input Pseudo-Doherty Load Modulated Balanced Amplifier using Signal-Flow-Based Phase Alignment Design}

\author{Pingzhu~Gong,~\IEEEmembership{Student Member,~IEEE},
Jiachen~Guo,~\IEEEmembership{Student Member,~IEEE},
Niteesh~Bharadwaj~Vangipurapu,~\IEEEmembership{Student Member,~IEEE}
        and Kenle~Chen,~\IEEEmembership{Senior Member,~IEEE}
\thanks{Manuscript received February xx, 2024; revised April xx, 2024; Date of publication April xx, 2024; This work was supported in part by the National Science Foundation under award no.~1914875 and 2218808.}
\thanks{The authors are with the Department
of Electrical and Computer Engineering, University of Central Florida, Orlando, FL, 32816, USA (e-mail: pingzhu.gong@ucf.edu and kenle.chen@ucf.edu).}
\thanks{This article was presented at the IEEE MTT-S International Microwave Symposium (IMS 2024), Washington, DC, USA, June 16–21, 2024.}

\thanks{Color versions of one or more of the figures in this paper are available online at https://ieeexplore.ieee.org.
}
\thanks{Digital Object Identifier 10.1109/LMWT.2024.3390585
}

\thanks{all rights of this paper under copyright that may exist in and to: (a) the Work, including any revised or expanded derivative works submitted to the IEEE by the undersigned based on the Work; and (b) any associated written or multimedia components or other enhancements accompanying the
Work belong to The Institute of Electrical and Electronics Engineers, Incorporated (the "IEEE").
}}

\markboth{IEEE MICROWAVE AND WIRELESS TECHNOLOGY LETTERS,~Vol.~xx, No.~x, JUNE~2024}%
{Shell \MakeLowercase{\textit{et al.}}: A Sample Article Using IEEEtran.cls for IEEE Journals}

\maketitle
\begin{abstract}
This paper reports a first-ever decade-bandwidth pseudo-Doherty load-modulated balanced amplifier (PD-LMBA), designed for emerging 4G/5G communications and multi-band operations. By revisiting the LMBA theory using the signal-flow graph, a frequency-agnostic phase-alignment condition is found that is critical for ensuring intrinsically broadband load modulation behavior. This unique design methodology enables, for the first time, the independent optimization of broadband balanced amplifier (BA, as the peaking) and control amplifier (CA, as the carrier), thus fundamentally addressing the longstanding limits imposed on the design of wideband load-modulated power amplifiers (PAs). To prove the proposed concept, an ultra-wideband RF-input PD-LMBA is designed and developed using GaN technology covering the frequency range from $0.2$ to $2$ GHz. Experimental results demonstrate an efficiency of $51\%$ to $72\%$ for peak output power and $44\%$ to $62\%$ for $10$-dB OBO, respectively.

\end{abstract}


\section{Introduction}
\IEEEPARstart{T}{o} meet the ever-increasing demands for higher data rates and transmission capacity, sophisticated modulation schemes are employed, resulting in signals with large amplitude variations characterized by peak-to-average power ratio (PAPR). Consequently, power amplifiers (PAs), as the most power-hungry units in wireless communication systems, suffer from significant efficiency degradation. Moreover, given the proliferation of the wireless spectrum, PAs need to operate over a wide frequency range.

\begin{figure}[t]
\centering
\includegraphics[width=8.3cm]{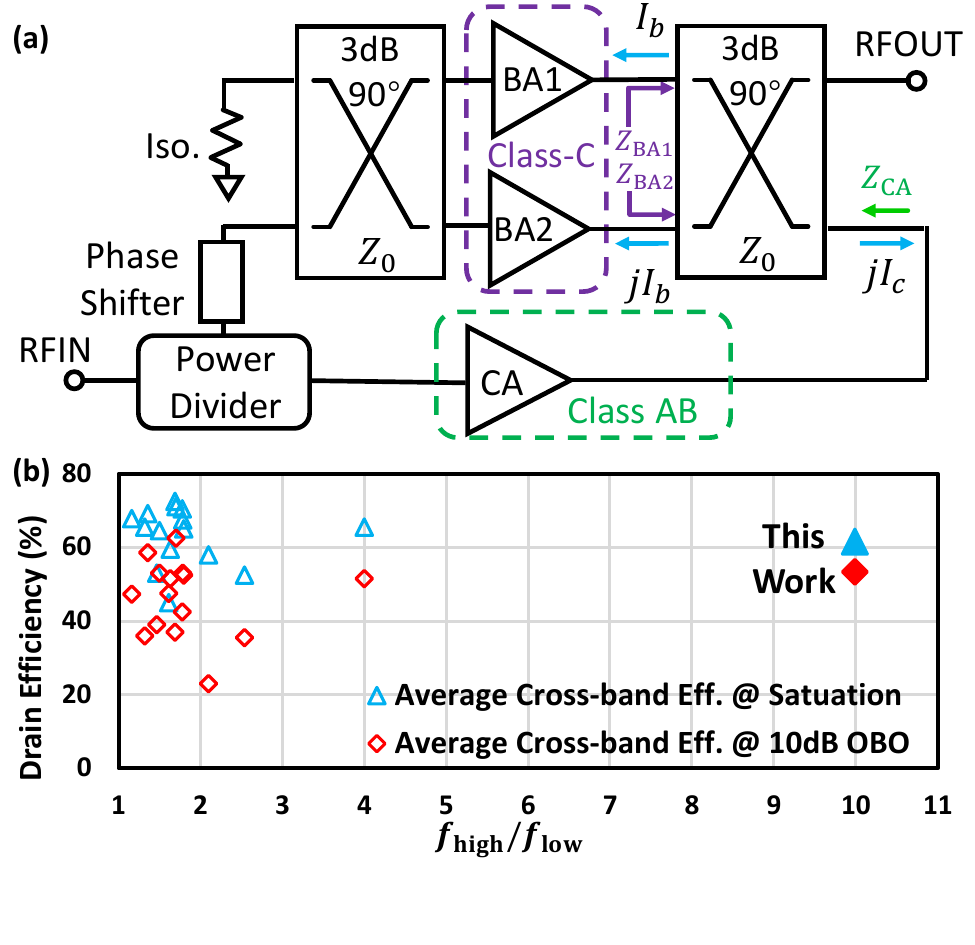}
\vspace{-1cm}
\caption{(a) General circuit schematic of PD-LMBA; (b) Comparison with state-of-the-art DPAs/LMBAs. ($f_{high}$/$f_{low}$ refers to the ratio between the upper and lower boundaries of frequency range. Average cross-band efficiency is defined as the average of maximum and minimum efficiency at a specific output power level.) }
\label{fig:overview}
\vspace{-0.7cm}
\end{figure}

To address the inefficiency in transmission of signals with high PAPR, load modulation (LM) has emerged as an effective solution for PAs. Doherty power amplifier (DPA), as a representative of load modulation, has been extensively studied \cite{DPA_roberto_2018}, \cite{Zhu_2021_DPA} and widely adopted in base station applications. However, DPA is limited in providing a wide bandwidth due to the intrinsic constraints of the impedance inverter. On the other hand, a recently introduced PA architecture, load-modulated balanced amplifier (LMBA) \cite{LMBA_MWCL2016}, has been demonstrated to provide both wide bandwidth and extended output power back-off (OBO) range. By injecting an additional signal into the isolation port of the output quadrature coupler of the balanced amplifier (BA) through another control amplifier (CA), the efficiency of BA is enhanced by load modulation. Furthermore, LMBA inherits the wideband nature of BA, offering a significant advantage over DPA. Developed from the original LMBA, a re-engineered LMBA mode is proposed, named as pseudo-Doherty load-modulated balanced amplifier (PD-LMBA) \cite{PDLMBA} or sequential load-modulated balanced amplifier (SLMBA) \cite{SLMBA}. In PD-LMBA/SLMBA, the CA is set as the carrier device and the BA as peaking as shown in Fig.~\ref{fig:overview}(a), enabling $>10$ dB of OBO range and ultra-wide bandwidth up to dual octaves ($4:1$) \cite{ALMBA}.

While LMBA is proven to be broadband in extensive experiments\cite{jiachen_HALMBA_TMTT2024,jiachen_1D_IMS2023}, no existing theory rigorously explains the consistency of load modulation across all in-band frequencies. By constructing the full signal-flow graph of LMBA, this article demonstrates that LMBA can consistently exhibit broadband load modulation behavior when a specific CA-BA phase-alignment condition is met. The proposed theory and method are validated through the development of a PD-LMBA prototype. The bandwidth and efficiency of the PD-LMBA prototype are then compared with DPAs and LMBAs reported in \cite{DPA_roberto_2018,Zhu_2021_DPA,PDLMBA,ALMBA,1D_LMBA,chenhao_SLMBA_2023,Pedro_dual_input_SLMBA_2023,chenyu_DPA_2023,xiaowei_three_stage_DPA_2019,jiachen_LMDBA_wamicon2023}, as illustrated in Fig.~\ref{fig:overview}(b). Designed using the proposed theory, the bandwidth of the PD-LMBA prototype is broadened unprecedentedly to a decade, significantly outperforming the state-of-the-art.

\section{PHASE-ALIGNMENT ANALYSIS BASED ON SIGNAL-FLOW GRAPH}

\begin{figure}[t]
\centering
\includegraphics[width=8.9cm]{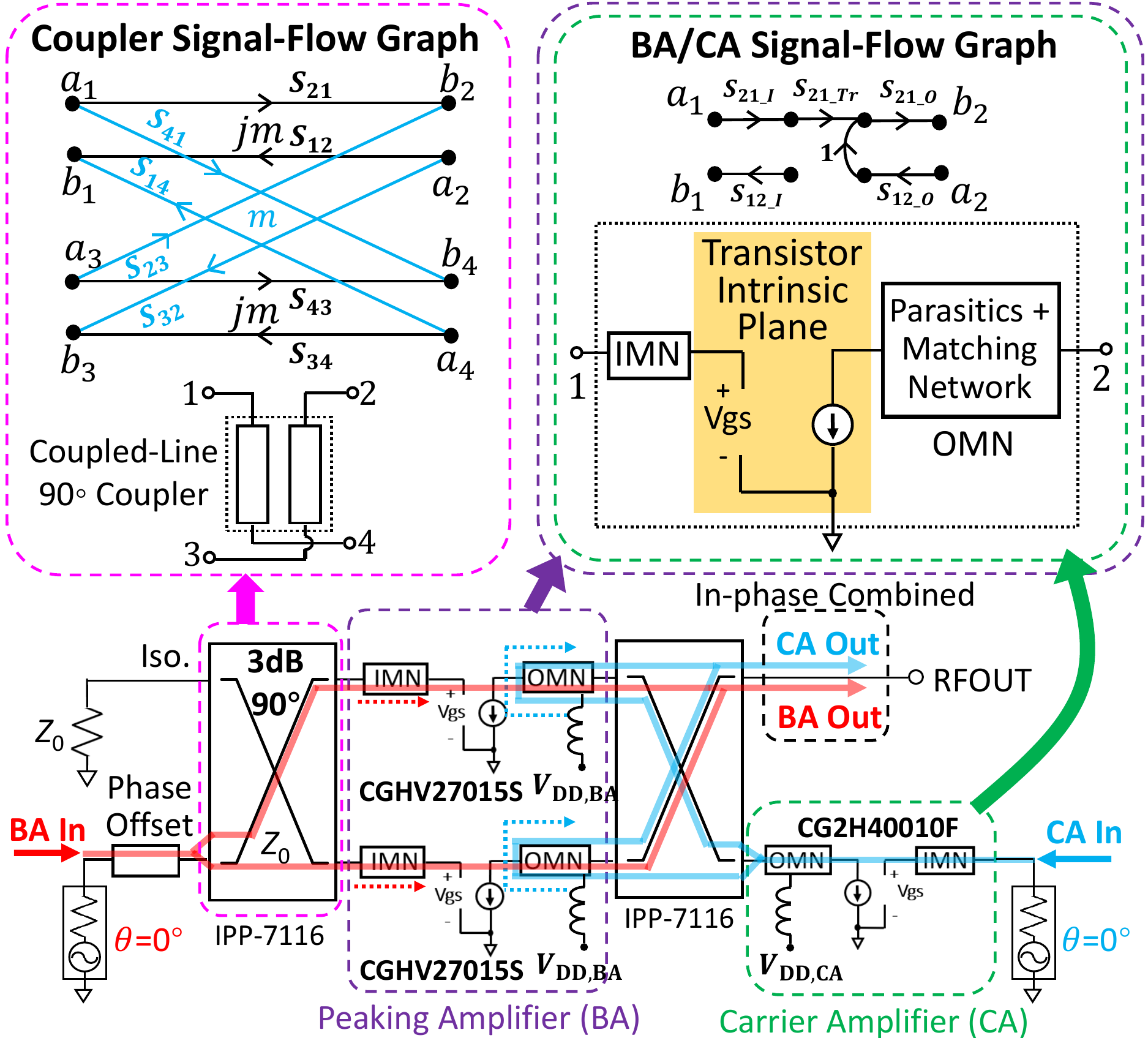}
\vspace{-0.7cm}
\caption{Proposed signal-flow graph for the wideband PD-LMBA with circuit implementation using wideband coupled line couplers and GaN transistors}
\label{fig:SFG}
\vspace{-0.5cm}
\end{figure}

The RF-input LMBA described in Fig.~\ref{fig:overview} consists of a balanced amplifier and a control amplifier. In the LMBA theory, load modulation is described by \cite{LMBA_MWCL2016,PDLMBA}: 
  \begin{align}
 &Z_{\mathrm{BA1}}=Z_{\mathrm{BA2}}=Z_{\mathrm{0}}(1+\frac{\sqrt{2}I_c e^{j\theta}}{I_b}).
\label{eq:Z_BA}
\end{align}
where $I_b$ and $I_c$ are the magnitude of BA and CA currents, respectively, and $\theta$ is the phase of the control path. Note that the classical LMBA theory indicated by Eq.~\eqref{eq:Z_BA} does not involve any frequency dependence. In contrast, our proposed LMBA theory intrinsically accounts for frequency-dependent variations using the signal-flow graph. This analysis proves the consistent load modulation behavior of LMBA across all in-band frequencies that solidly explains its wideband nature.

\subsection{LMBA Revisited using Signal-Flow Graph}

The $S$-matrix of wideband coupled-line coupler is given by:
\begin{equation}
 \begin{bmatrix}
   b_{1} \\
   b_{2} \\
    b_{3}\\
    b_{4} 
  \end{bmatrix}
=
 \begin{bmatrix}
  0&jm&0&m\\
    jm&0&m&0\\
   0&m&0&jm \\
   m&0&jm&0
  \end{bmatrix}
  \begin{bmatrix}
   a_{1} \\
   a_{2} \\
    a_{3}\\
    a_{4} 
  \end{bmatrix}
  \label{eq:S_coupler}
\end{equation}
where $m=\frac{1}{\sqrt{2}} {e^{j\theta(\omega)}}$.  While the phase of $m$ is denoted by $\theta(\omega)$ representing its frequency dependence, the relationship of $S_{\mathrm{21}} = j S_{\mathrm{41}}$ consistently holds true. This implies that a $90^\circ$ phase shift between the thru and coupled ports can be maintained across the entire frequency range of coupler.

The $S$-matrix of the transistor in a PA is expressed as:
\begin{equation}
[S_{\mathrm{Tr}}] = \begin{bmatrix} 0 & 0 \\ S_{\mathrm{21,Tr}} & 1 
\label{eq:S_tr}
\end{bmatrix}
\end{equation}
The transistor is modeled as an ideal voltage-controlled current source with infinite reverse isolation, and it is assumed to be matched at the input. Furthermore, the $S$-matrix of input/output matching network of a PA is represented as: 
\begin{equation}
[S_{\mathrm{I/O}}] = \begin{bmatrix} 0 & S_{\mathrm{12,I/O}} \\ S_{\mathrm{21,I/O}} & 0 \end{bmatrix}
\label{eq:S_MN}
\end{equation} Both the input matching network (IMN) and output matching network (OMN) are assumed to be matched and reciprocal. 

Based on the $S$-matrix of \eqref{eq:S_coupler}-\eqref{eq:S_MN} and Fig.~\ref{fig:overview}, the signal-flow graph of generic LMBA is depicted in Fig.~\ref{fig:SFG}, which can be viewed as the combination of two couplers and three sub-PAs with IMN and OMN included. Note that the device parasitics is considered as part of the OMN in \eqref{eq:S_tr}. The $S$-matrix of the phase shifter can also be modeled by \eqref{eq:S_MN}.

By applying Mason's rules\cite{Pozar_mircrowave} to the signal-flow graph of LMBA in Fig.~\ref{fig:SFG}, the output waves ($b_\mathrm{out,BA}$, $b_\mathrm{out,CA}$) induced by the BA and CA inputs ($a_\mathrm{BA}$, $a_\mathrm{CA}$) can be expressed as: \begin{align}
\begin{split}
b_{\mathrm{out,BA}}={a_{\mathrm{BA}}}\cdot {2jm^{2}} {S_{\mathrm{21,PHS}}} {S_{\mathrm{21,BI}}} {S_{\mathrm{21,BTr}}} {S_{\mathrm{21,BO}}},
\\
b_{\mathrm{out,CA}}={a_{\mathrm{CA}}}\cdot {2jm^{2}}  {S_{\mathrm{21,CI}}} {S_{\mathrm{21,CTr}}} {S_{\mathrm{21,CO}}} {S_{\mathrm{21,BO}}}^2
\label{eq:bout_BA/CA}
\end{split}
\end{align}where ${S_{\mathrm{21,PHS}}}$ stands for the ${S_{\mathrm{21}}}$ of the phase shifter, and subscripts B and C refer to BA and CA, respectively (e.g., ${S_{\mathrm{21,BI}}}$ denotes the ${S_{\mathrm{21}}}$ of the BA IMN). The total output wave is a combination of BA and CA, i.e., $b_{\mathrm{out}}=b_{\mathrm{out,BA}}+b_{\mathrm{out,CA}}$. In Fig.~\ref{fig:SFG}, the signal paths of BA (in red) and CA (in blue) are identified and visualized based on \eqref{eq:bout_BA/CA}, similar to \cite{1D_LMBA}. The common factor ${jm^{2}}$ implies that both the BA signal and CA signal pass through the coupler twice, resulting in the same phase delay, even though the phase of $m$ is a function of frequency. The shared factor $S_{\mathrm{21,BO}}$ indicates that the BA signal passes through the BA OMN once, while the CA signal traverses the BA OMN twice. 

\subsection{Frequency-Agnostic Phase Alignment for PD-LMBA}

Given the ideal load modulation behavior of PD-LMBA, i.e., $\theta=0^\circ$ in \eqref{eq:Z_BA}, BA and CA signals should be in-phase combined at the output. Eliminating the common factors in \eqref{eq:bout_BA/CA}, the following condition needs to be satisfied:
\begin{equation}
\begin{split}
&\angle{S_{\mathrm{21,PHS}}} {S_{\mathrm{21,BI}}} {S_{\mathrm{21,BTr}}}=\angle{S_{\mathrm{21,CI}}} {S_{\mathrm{21,CTr}}} {S_{\mathrm{21,CO}}} {S_{\mathrm{21,BO}}}
\label{eq:phase}
\end{split}
\end{equation}Note that Eqs.~\eqref{eq:bout_BA/CA} and \eqref{eq:phase} contain the frequency dependence of all building blocks. If we assume the similar behavior of IMNs and transistors between BA and CA ($\angle S_\mathrm{21,BI}S_\mathrm{21,BTr}\approx\angle S_\mathrm{21,CI}S_\mathrm{21,CTr}$), the input phase shifter only needs to offset the combined phase of CA and BA OMNs ($\angle{S_\mathrm{21,CO}S_\mathrm{21,BO}}=\angle{S_\mathrm{21,PHS}}$), so as to realize an inherently wideband PD-LMBA. 



\begin{figure}[t]
\centering
\includegraphics[width=8cm]{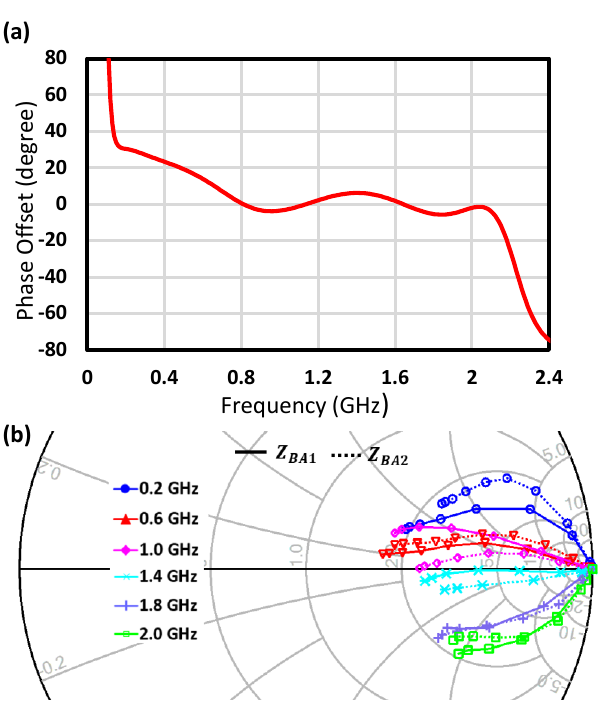}
\vspace{-0.5cm}
\caption{(a) BA and CA signal path phase offset based on (\ref{eq:phase}) at different frequencies. (b) BA Intrinsic load impedance trajectories}
\label{fig:Fig3}
\vspace{-0.5cm}
\end{figure}

\begin{figure}[t]
\centering
\includegraphics[width=8.5cm]{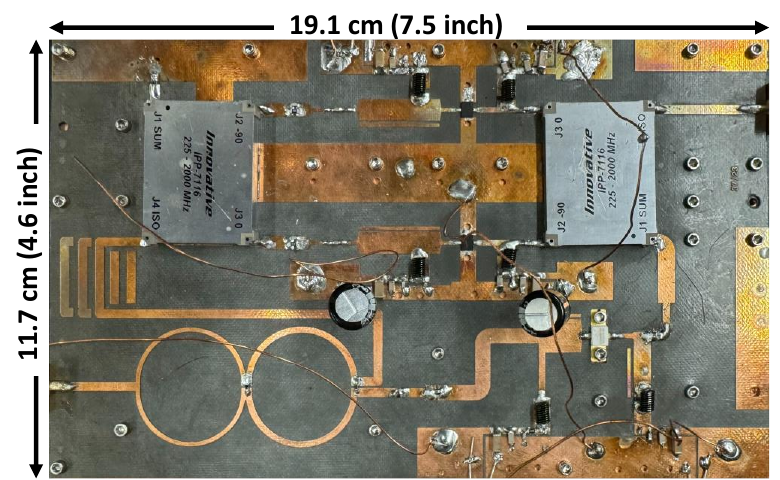}
\vspace{-0.2cm}
\caption{Fabricated PD-LMBA prototype.}
\label{fig:Fab}
\vspace{-0.6cm}
\end{figure}

\section{Design of Decade-Bandwidth PD-LMBA}

Based on the proposed theory, a $10$-W GaN transistor (Wolfspeed CG2H40010F) is employed for realizing the CA, while two $15$-W GaN transistors (Wolfspeed CGHV27015S) are used for the BA in the PD-LMBA prototype. The realized circuit schematic is shown in Fig.~\ref{fig:SFG}. The target OBO is set to $10$ dB to handle high-PAPR signals, and the target frequency range is from $0.2$ to $2$ GHz. Two identical $50$-$\Omega$ wideband couplers (IPP-7116, Innovative Power Products) are employed at the input and output of the BA, which provides the same phase delay for both BA and CA.

Device parasitics are extracted using the method reported in \cite{pingzhu_MS_thesis}, with intrinsic nodes of the transistor provided by the model. Subsequently, these device parasitics are utilized as part of OMN. Multisection-transformer-based matching network are employed to realize BA and CA IMNs, as shown in Fig.~\ref{fig:Fab}. Three RC-based networks (with R = 200 $\Omega$ and C = 10 pF) are incorporated into the IMNs to stabilize the PA. For CA OMN, a shunt open stub is added right after the transistor drain, along with two transmission lines in series, to collaborate with CA parasitics and establish the broadband multisection 40 $\Omega$ - 50 $\Omega$ matching \cite{DPA_roberto_2018}. Additionally, a short transmission line is introduced at the BA output to resonate with BA parasitics and provide 50 $\Omega$ - 50 $\Omega$ matching. The IMNs and OMNs for both BAs and CAs are designed with the minimum number of shunt stubs to minimize phase dispersion, simplifying the phase-alignment requirement described by (\ref{eq:phase}). This allows us to add a transmission-line based phase shifter at the BA input to properly align the phase of BA and CA path. The phase offset between the BA and CA signal paths is plotted in Fig.~\ref{fig:Fig3}(a), revealing that the phase difference in the proposed PD-LMBA is below $30$ degrees from $0.2$ GHz to $2$ GHz. Fig.~\ref{fig:Fig3}(b) displays the intrinsic load trajectories of BAs, indicating a desired load modulation behavior across all in-band frequencies.


\begin{figure}[t]
\centering
\includegraphics[width=8.8cm]{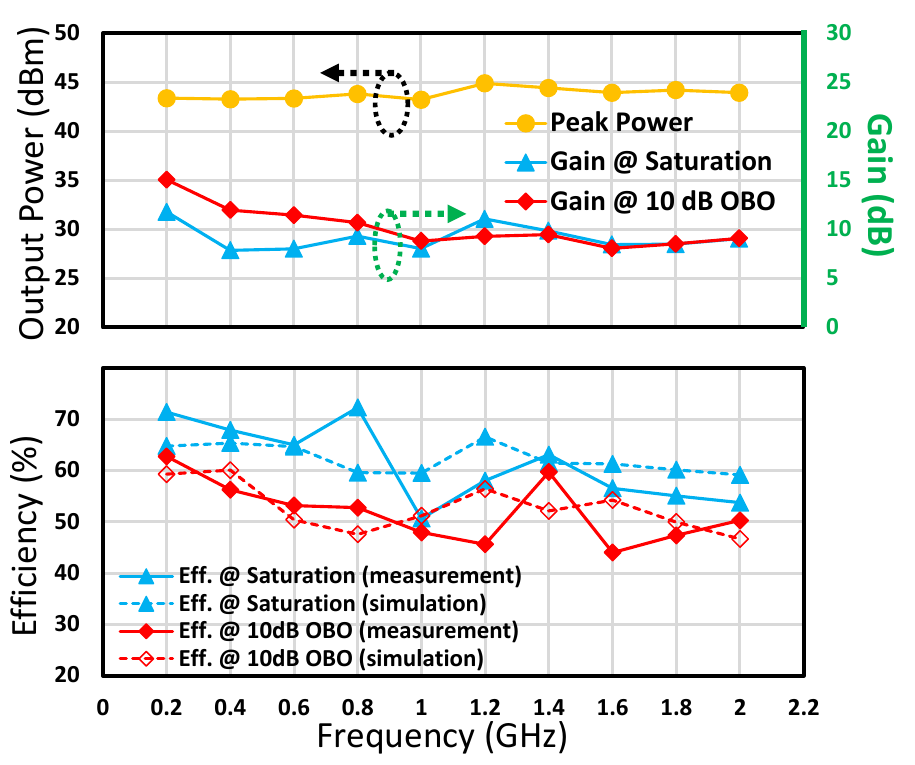}
\vspace{-0.9cm}
\caption{Measured peak output power, gain and efficiency at various OBO levels from $0.2$ to $2$ GHz.}
\label{fig:figure5_Power_EFF}
\vspace{-0.4cm}
\end{figure}

\begin{figure}[t]
\centering
\includegraphics[width=9cm]{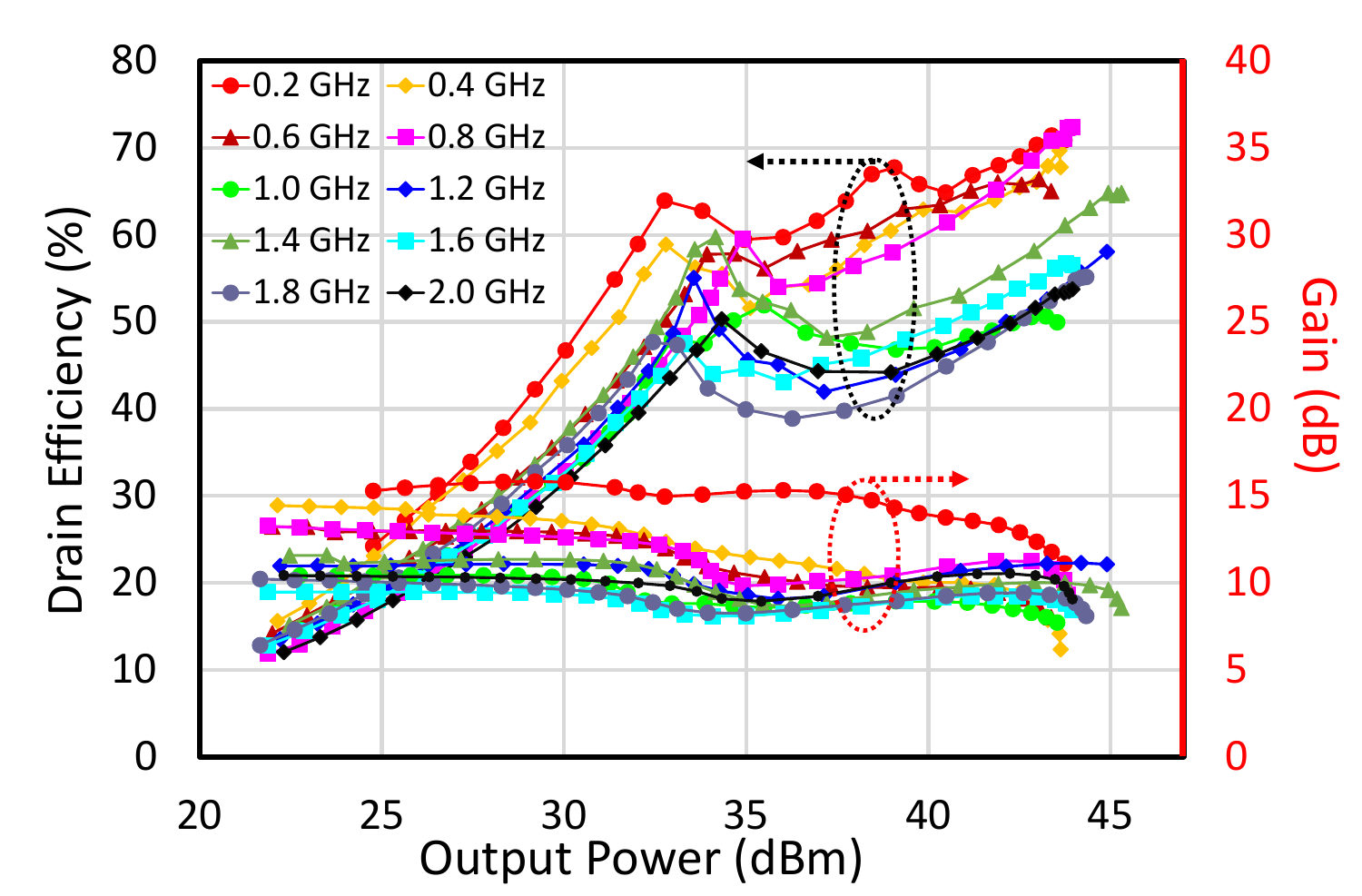}
\vspace{-0.8cm}
\caption{Power-swept measurement of efficiency and gain from $0.2$ to $2$ GHz.}
\label{fig:EffGain_vs_Pout}
\vspace{-0.53cm}
\end{figure}

\newcommand{\tabincell}[2]{\begin{tabular}{@{}#1@{}}#2\end{tabular}}

\section{Implementation and Experimental Results}
The PA is implemented on a $31$-mil thick Rogers Duroid-5880 PCB board with a dielectric constant of $2.2$, as shown in Fig.~\ref{fig:Fab}. The  CA is biased in Class-AB with a $V_\mathrm{DD,CA}$ around $12$ V. The BA is biased in Class-C with $50$-V $V_\mathrm{DD,BA}$. Bias voltages are adjusted to improve the PA performance at different frequencies. The prototype is tested using both continuous-wave (CW) and modulated signals.

\subsection{Continuous-Wave Measurement} 

The PD-LMBA prototype is evaluated using a single-tone signal across the frequency range of $0.2$ to $2$ GHz at various power levels. The frequency response is shown in Fig.~\ref{fig:figure5_Power_EFF}. A peak output power ranging from $43-45$ dBm is observed across the entire bandwidth, along with a gain of $9-15$ dB at different  OBO levels. The corresponding measured peak efficiency falls within the range of $51-72\%$, and the efficiencies at $10$ dB OBOs are in the range of $44-62\%$. The power-dependent gain and efficiency profiles at various frequencies are shown in Fig.~\ref{fig:EffGain_vs_Pout}, which indicates a strong efficiency enhancement across different power levels. The $V_\mathrm{GS,BA1}$ and $V_\mathrm{GS,BA2}$ are set unequal at some frequencies to further improve the back-off efficiency between the peak power and $10$-OBO \cite{CM_LMBA}, e.g., $0.2$ GHz in Fig.~\ref{fig:EffGain_vs_Pout}. 



\begin{figure}[t]
\centering
\includegraphics[width=8.9cm]{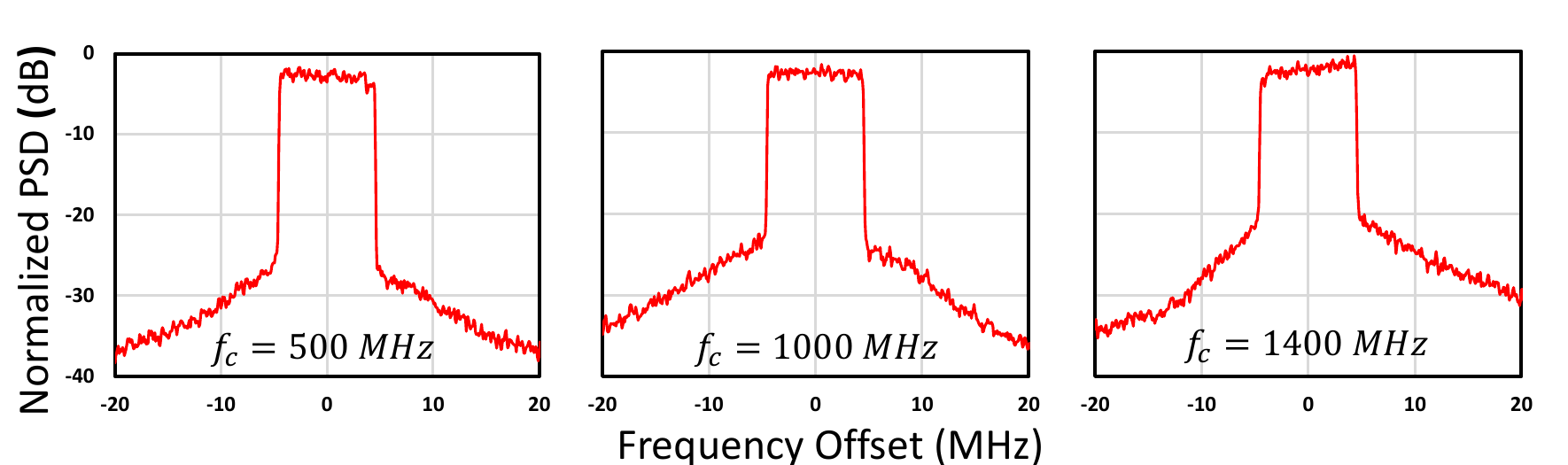}
\caption{Output spectrum from modulated measurement using a $10$-MHz $9.5$-dB-PAPR LTE signal centered at $500$, $1000$ and $1400$ MHz.}
\label{fig:output spectrum}
\vspace{-0.5cm}
\end{figure}

\subsection{Modulated Measurements} 
A $10$-MHz-bandwidth LTE signal with a PAPR of $9.5$ dB is used to perform modulated measurement. The modulated signals are generated and analyzed using Keysight PXIe vector transceiver (VXT M9421). The average power of the modulated signals is $34$ dBm, and the output spectrum is shown in Fig.~\ref{fig:output spectrum} at $500$, $1000$, and $1400$ MHz, with average efficiency of $52\%$, $48\%$, and $49\%,$ respectively. The measured adjacent channel leakage ratios (ACLR) at these frequencies are higher than $22$ dB, and no digital predistortion (DPD) is performed.


\section{Conclusion}
This paper presents a comprehensive analysis of LMBA from a signal-flow perspective. By constructing the full signal-flow graph of LMBA, it is, for the first time, demonstrated that the load modulation behavior of LMBA can be sustained across all in-band frequencies when a specific BA-CA phase alignment condition is met. This finding not only provides a rigorous explanation of the wideband nature of LMBA but also introduces a novel design methodology for realizing ultra-wideband LMBAs. Moreover, the signal-flow approach offers the opportunity to analyze the behavior of other load-modulated PA topologies from a new perspective. To prove the proposed concept, an ultra-wideband RF-input PD-LMBA is designed and developed using GaN technology covering the frequency range from $0.2$ to $2$ GHz. Experimental results demonstrate an efficiency of $51\%$ to $72\%$ for peak output power and $44\%$ to $62\%$ for $10$-dB OBO, respectively. By leveraging the BA-CA phase alignment condition identified in this study, the bandwidth of the PD-LMBA prototype has been extended to an unprecedented decade-wide range, significantly outperforming the state-of-the-art.

\section{Acknowledgement}
The authors would like to thank Prof. Patrick Roblin of The Ohio State University for his insightful signal-flow graph of the balanced amplifier from his lecture slides. The authors also want to thank Ryan Baker of Wolfspeed (now MACOM) for providing transistor samples.
\bibliographystyle{IEEEtran}
\bibliography{References.bib}

\begin{thebibliography}{10}
\providecommand{\url}[1]{#1}
\csname url@samestyle\endcsname
\providecommand{\newblock}{\relax}
\providecommand{\bibinfo}[2]{#2}
\providecommand{\BIBentrySTDinterwordspacing}{\spaceskip=0pt\relax}
\providecommand{\BIBentryALTinterwordstretchfactor}{4}
\providecommand{\BIBentryALTinterwordspacing}{\spaceskip=\fontdimen2\font plus
\BIBentryALTinterwordstretchfactor\fontdimen3\font minus \fontdimen4\font\relax}
\providecommand{\BIBforeignlanguage}[2]{{%
\expandafter\ifx\csname l@#1\endcsname\relax
\typeout{** WARNING: IEEEtran.bst: No hyphenation pattern has been}%
\typeout{** loaded for the language `#1'. Using the pattern for}%
\typeout{** the default language instead.}%
\else
\language=\csname l@#1\endcsname
\fi
#2}}
\providecommand{\BIBdecl}{\relax}
\BIBdecl

\bibitem{DPA_roberto_2018}
J.~J. Moreno~Rubio, V.~Camarchia, M.~Pirola, and R.~Quaglia, ``Design of an 87\% fractional bandwidth doherty power amplifier supported by a simplified bandwidth estimation method,'' \emph{IEEE Transactions on Microwave Theory and Techniques}, vol.~66, no.~3, pp. 1319--1327, 2018.

\bibitem{Zhu_2021_DPA}
Y.~Xu, J.~Pang, X.~Wang, and A.~Zhu, ``Enhancing bandwidth and back-off range of doherty power amplifier with modified load modulation network,'' \emph{IEEE Transactions on Microwave Theory and Techniques}, vol.~69, no.~4, pp. 2291--2303, 2021.

\bibitem{LMBA_MWCL2016}
D.~J. {Shepphard}, J.~{Powell}, and S.~C. {Cripps}, ``An efficient broadband reconfigurable power amplifier using active load modulation,'' \emph{IEEE Microw. Wireless Compon. Lett.}, vol.~26, no.~6, pp. 443--445, June 2016.

\bibitem{PDLMBA}
Y.~{Cao} and K.~{Chen}, ``Pseudo-{Doherty} load-modulated balanced amplifier with wide bandwidth and extended power back-off range,'' \emph{IEEE Trans. Microw. Theory Techn.}, vol.~68, no.~7, pp. 3172--3183, 2020.

\bibitem{SLMBA}
J.~{Pang}, Y.~{Li}, M.~{Li}, Y.~{Zhang}, X.~Y. {Zhou}, Z.~{Dai}, and A.~{Zhu}, ``Analysis and design of highly efficient wideband {RF}-input sequential load modulated balanced power amplifier,'' \emph{IEEE Trans. Microw. Theory Techn.}, vol.~68, no.~5, pp. 1741--1753, 2020.

\bibitem{ALMBA}
Y.~{Cao}, H.~{Lyu}, and K.~{Chen}, ``Asymmetrical load modulated balanced amplifier with continuum of modulation ratio and dual-octave bandwidth,'' \emph{IEEE Trans. Microw. Theory Techn.}, vol.~69, no.~1, pp. 682--696, 2021.

\bibitem{jiachen_HALMBA_TMTT2024}
J.~Guo, Y.~Cao, and K.~Chen, ``Linear hybrid asymmetrical load-modulated balanced amplifier with multiband reconfigurability and antenna-vswr resilience,'' \emph{IEEE Transactions on Microwave Theory and Techniques}, pp. 1--14, 2024.

\bibitem{jiachen_1D_IMS2023}
J.~Guo and K.~Chen, ``Reconfigurable hybrid asymmetrical load modulated balanced amplifier with high linearity, wide bandwidth, and load insensitivity,'' in \emph{2023 IEEE/MTT-S International Microwave Symposium - IMS 2023}, 2023, pp. 462--465.

\bibitem{1D_LMBA}
J.~Guo, Y.~Cao, and K.~Chen, ``{1-D} reconfigurable pseudo-doherty load modulated balanced amplifier with intrinsic {VSWR} resilience across wide bandwidth,'' \emph{IEEE Transactions on Microwave Theory and Techniques}, vol.~71, no.~6, pp. 2465--2478, 2023.

\bibitem{chenhao_SLMBA_2023}
C.~Chu, J.~Pang, R.~Darraji, S.~K. Dhar, T.~Sharma, and A.~Zhu, ``Broadband sequential load modulated balanced amplifier with extended design space using second harmonic manipulation,'' \emph{IEEE Transactions on Microwave Theory and Techniques}, vol.~71, no.~5, pp. 1990--2003, 2023.

\bibitem{Pedro_dual_input_SLMBA_2023}
C.~Belchior, L.~C. Nunes, P.~M. Cabral, and J.~C. Pedro, ``Sequential {LMBA} design technique for improved bandwidth considering the balanced amplifiers off-state impedance,'' \emph{IEEE Transactions on Microwave Theory and Techniques}, vol.~71, no.~8, pp. 3629--3643, 2023.

\bibitem{chenyu_DPA_2023}
C.~Liang, P.~Roblin, Y.~Hahn, J.~I. Martinez-Lopez, H.-C. Chang, and V.~Chen, ``Single-input broadband hybrid doherty power amplifiers design relying on a phase sliding-mode of the load modulation scheme,'' \emph{IEEE Transactions on Microwave Theory and Techniques}, vol.~71, no.~4, pp. 1550--1562, 2023.

\bibitem{xiaowei_three_stage_DPA_2019}
J.~Xia, W.~Chen, F.~Meng, C.~Yu, and X.~Zhu, ``Improved three-stage doherty amplifier design with impedance compensation in load combiner for broadband applications,'' \emph{IEEE Transactions on Microwave Theory and Techniques}, vol.~67, no.~2, pp. 778--786, 2019.

\bibitem{jiachen_LMDBA_wamicon2023}
J.~Guo and K.~Chen, ``Load-modulated double-balanced amplifier with quasi-isolation to load,'' in \emph{2023 IEEE Wireless and Microwave Technology Conference (WAMICON)}, 2023, pp. 144--147.

\bibitem{Pozar_mircrowave}
\BIBentryALTinterwordspacing
D.~M. Pozar, ``{Microwave Engineering; 3rd ed.}'' Hoboken, NJ, 2005. [Online]. Available: \url{https://cds.cern.ch/record/882338}
\BIBentrySTDinterwordspacing

\bibitem{pingzhu_MS_thesis}
P.~{Gong}, ``Design of a broadband doherty power amplifier with a graphical user interface tool,'' Master's Thesis, Ohio State University, \url{http://rave.ohiolink.edu/etdc/view?acc\_num=osu1658158564568844}, 2022.

\bibitem{CM_LMBA}
Y.~Cao, H.~Lyu, and K.~Chen, ``Continuous-mode hybrid asymmetrical load-modulated balanced amplifier with three-way modulation and multi-band reconfigurability,'' \emph{IEEE Transactions on Circuits and Systems I: Regular Papers}, vol.~69, no.~3, pp. 1077--1090, 2022.

\end{thebibliography}
\end{document}